\documentclass{rspublic}
 
\usepackage{epsf}
\usepackage{longtable}
\usepackage{natbib}
\usepackage{epstopdf}
\usepackage{graphicx}

\def \apj   {ApJ}
\def \apjs   {ApJS}
\def \apjl   {ApJL}
\def \mnras   {MNRAS}
\def \solphys   {SolPhys}
\def \pasp   {PASP}
\def \araa   {ARA\&A}
\def \aap   {A\&A}
\def \aapr   {A\&ARv}
\def \physscr   {PhyS}
\def \aj   {AJ}
\def \ssr   {SSRv}
\def \nat   {Nature}
\def \jgr   {JGR}
\def \aaps   {A\&AS}
\def \grl   {GRL}
\def \pasj   {PASJ}

\def \caii   {Ca\,{\sc ii}}
\def \mgii   {Mg\,{\sc ii}}
\def \siiv   {Si\,{\sc iv}}
\def \civ   {C\,{\sc iv}}

\def \rossn {$R_{0}$}
\def \rossnsat {$R_{0_{sat}}$}
\def \lx {L$_{\rm X}$}
\def \prot {P$_{rot}$}
\def \lbol {L$_{bol}$}
\def \fx {F$_{\rm X}$}
\def \rx {\lx/\lbol}
\def \msol {$M_{\odot}$}
\def \phiv {$\phi_{\rm V}$}
\def \bv {$\langle |B_{\rm V}| \rangle$}
\def \bi {$\langle |B_{\rm I}| \rangle$}

\def \mstar {M$_{\star}$}

\def \rhk {$R'_{HK}$}
\def \teff {$T_{eff}$}
\def \mfd {$|\phi|$}

\def \einstein   {{\em EINSTEIN}}
\def \chandra   {{\em Chandra}}
\def \xmm   {{\em XMM-Newton}}

\begin{document} 
 
\title[Stellar Activity and Coronal Heating]{Stellar Activity and Coronal Heating: an overview of recent results} 
 
\author[P.~Testa, S.H.~Saar, J.J.~Drake]{Paola Testa$^{1,*}$, Steven H. Saar$^{1}$, Jeremy J. Drake$^{1}$} 
 
\affiliation{$^1$Harvard-Smithsonian Center for Astrophysics, 60 Garden st., Cambridge, MA 02138, USA.
$^*$ Correspondence to: ptesta@cfa.harvard.edu} 
 
\maketitle 
 
\begin{abstract}{solar corona, magnetic activity, X-ray activity, chromospheric activity, solar-stellar connection} 
Observations of the coronae of the Sun and of solar-like stars provide complementary information to advance our understanding of stellar magnetic activity, and of the processes leading to the heating of their outer atmospheres. While solar observations allow us to study the corona at high spatial and temporal resolution, the study of stellar coronae allows us to probe stellar activity over a wide range of ages and stellar parameters. Stellar studies therefore provide us with additional tools for understanding coronal heating processes, as well as the long-term evolution of solar X-ray activity. We discuss how recent studies of stellar magnetic fields and coronae contribute to our understanding of the phenomenon of activity and coronal heating in late-type stars.
\end{abstract} 
 
\section{Introduction} 
\label{sec:intro}

The detailed physical processes that heat the outer atmosphere of the Sun and of solar-like stars to millions of degrees are still poorly understood, and remain a major open issue in astrophysics. Recently, significant progress has been made, both observationally and theoretically, in understanding coronal heating mechanisms (see also other papers of this issue on ``New approaches in coronal heating'' for reviews of different aspects of recent coronal heating research). 

Stellar magnetic activity is ubiquitous in solar-like stars, i.e., relatively unevolved stars with convective envelopes below their photospheres, and it includes a variety of phenomena such as starspots, activity cycles, hot outer atmospheres (comprised of chromosphere, transition region and corona), and magnetic breaking due to stellar winds \citep[see e.g., a review by][]{schrijver08}. Understanding magnetic activity is of fundamental importance for stellar astrophysics since it plays a crucial role in star and planet formation and evolution, controls stellar angular momentum loss, and drives orbital decay and evolution in close binary systems.

\begin{figure}[!ht]
\centering
\scalebox{.5}{\includegraphics{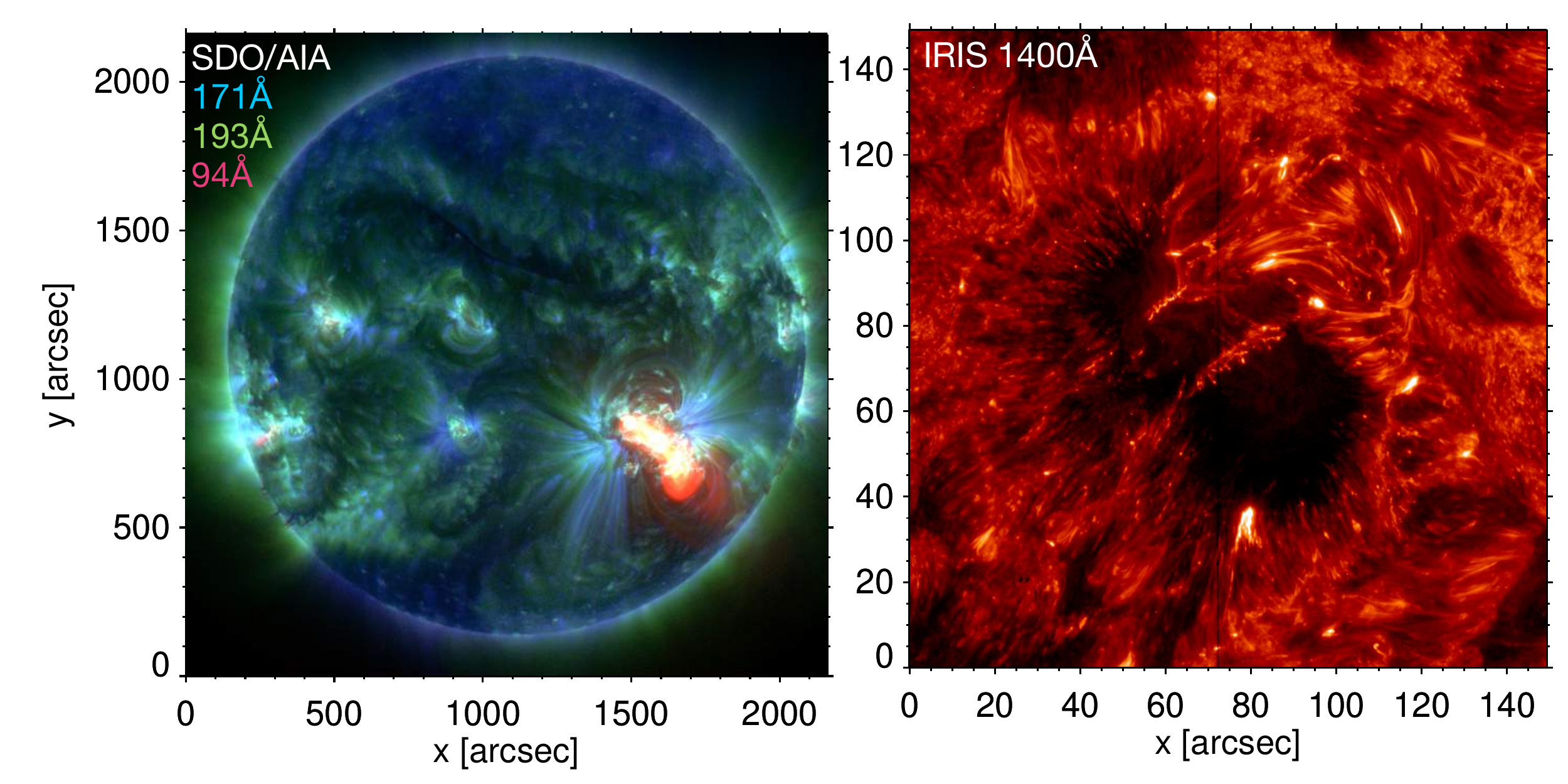}}
\caption{Example of recent imaging observations of the solar corona taken with the Atmospheric Imaging Assembly ({\em AIA}; \citealt{lemen12}) onboard the {\em Solar Dynamics Observatory} ({\em left}), and with the Interface Region Imaging Spectrograph (\citealt{depontieu14}; {\em right}). {\em AIA} provides full-disk images in narrow EUV passbands, at 12 sec cadence and $\sim 0.6$~arcsec/pix, and {\em IRIS} provides simultaneous spectra and images of the photosphere, chromosphere, transition region, and corona, with temporal resolution of seconds, $\sim 0.166$~arcsec/pix, and $\sim 1$~km s$^{-1}$ velocity resolution, over a field-of-view of up to 175 arcsec $\times$ 175 arcsec.
{\em Left:} three-color full disk image combining {\em AIA} observations in three narrow EUV passbands centered around 94\AA\ (red), 193\AA\ (green), and 171\AA\ (blue), respectively. {\em Right:} slit-jaw images in the {\em IRIS} FUV passband centered around 1400\AA, of active region 12192, which is the largest active region observed on the Sun since 1990. Both observations are taken on 2014-10-25 around 18UT.}
\label{fig:sun_obs}
\end{figure}

Solar observations provide us with a close up view of magnetic activity, allowing us to investigate the characteristics of the magnetic field and magnetized plasma at high spatial and temporal resolution. In Fig. \ref{fig:sun_obs} we show an example of recent solar observations at UV/EUV/X-ray wavelengths, able to resolve coronal structures down to $\sim 250$~km. 
Stellar observations, on the other hand, allow us to explore how the stellar activity depends on stellar parameters (e.g., stellar mass, age, chemical composition), and to test the validity and limitations of the solar-stellar analogy, and put tighter constraints on models of activity and coronal heating. 

In this paper we focus on the stellar perspective on magnetic activity, and in particular we discuss the insights that can be gained on coronal activity and coronal heating from stellar studies. 
By necessity, we will confine the discussion to a few selected topics that we believe have provided more significant advances in recent years, and in particular will review results based on observational studies. We refer the reader to earlier reviews for in-depth discussion of other aspects (including theoretical modeling) that are not included or only very briefly touched upon here (e.g., dynamo models, stellar chromospheric emission and cycles, starspots, activity in pre-main sequence stars, star-planet interactions and effects on stellar activity; e.g., \citealt{durney93,charbonneau10,linsky80,berdyugina05,strassmeier09,favata03,preibisch05,guedel09,testa10,poppenhaeger14}). 

In Section \ref{sec:1}, we briefly review the basic properties of magnetic activity in solar-like stars, focusing in particular on X-ray observations and coronal heating processes. We then discuss a few selected aspects of magnetic activity in some detail, including its correlation with stellar parameters and evolution with stellar age. In Section \ref{sec:2}, we highlight the recent results on activity cycles and magnetic grand minima, and in Section \ref{sec:3}, we discuss the chemical  abundance anomalies in stellar coronae and their possible diagnostic potential for coronal heating. Finally in Section \ref{sec:4},  we discuss the results we have highlighted and conclude with a series of key open questions.

 \section{Stellar magnetic activity, its correlation with stellar parameters, and its evolution} 
\label{sec:1}
 
The evidence of a hot outer atmosphere on the Sun was initially interpreted as a direct consequence of convection, via dissipation of acoustic modes generated by turbulent convection \citep[e.g.,][]{schwarzschild48,kuperus69}. This scenario came into question following the first X-ray observations of the Sun and other stars.
A rapid improvement of instrumental capabilities since the first X-ray solar observations \citep{friedman51} yielded X-ray images of the Sun's high-energy emission, from rocket experiments \citep[e.g.,][]{blake63,vaiana68} and Skylab \citep[e.g.,][]{vaiana73} in the 60's and 70's. These observations revealed the highly structured and inhomogeneous nature of the hot corona, and their relation to magnetic features \citep[e.g.,][]{vanspeybroeck70}.
Systematic X-ray observations of late-type stars with space observatories \citep[e.g.,][]{vaiana81}, revealing the wide range in X-ray luminosity of stars with similar spectral types, further demonstrated the inadequacy of acoustic heating in explaining the emission from stellar coronae.  Instead, magnetic heating seemed to be required \citep[e.g.][see also reviews by e.g., \citealt{haisch96,ulmschneider91}]{rosner80}. 

While progress in solar and stellar coronal observations has helped to develop and constrain models for coronal heating, the details of the physical processes converting magnetic energy into atmospheric heating are not yet well established. The most favored candidate heating mechanisms include dissipation of magnetohydrodynamic (Alfv$\acute{\rm e}$n) waves \citep[e.g.,][]{vanballegooijen11,vanballegooijen14}, and dissipation of magnetic stresses as in the ``nanoflare'' model, where random photospheric motions lead to braiding of magnetic field lines and subsequent magnetic reconnection \citep[e.g.,][]{parker88,priest02}. Most viable models predict energy release characterized by small spatial and temporal scales, typically below current resolution capabilities for the Sun, and constraining the models with observations presents significant challenges \citep[see reviews by e.g.,][]{klimchuk06,reale14}.

The main characteristics of X-ray emission in late-type stars can be summarized as follows \citep[see also e.g.,][for reviews]{guedel04rev,testa10}: (1) all unevolved late-type stars (F-M type) are X-ray sources \citep[e.g.][]{vaiana81,Schmitt04}, as are evolved late-type stars earlier than spectral type mid-K \citep{haisch92}; (2) they show a very wide range in X-ray emission level of several orders of magnitude, even among the same spectral types; (3) they appear to have a minimum X-ray surface flux, \fx, value ($\sim 10^{4}$ erg cm$^{-2}$ s$^{-1}$; \citealt[e.g.,][]{Schmitt04}); (4) their X-ray activity, as parametrized by the ratio \rx\ of the X-ray luminosity, \lx, to the total (bolometric) luminosity, \lbol, appears to saturate at a value of $\log ($\rx $)\sim -3$ (\citealt{vilhu84,vilhu87}; see sec.2.\ref{ssec:1.1} below for a discussion of the implications of some of these properties on dynamo action at work in late-type stars).
 
As mentioned above, X-ray emission is only one of the aspects of stellar magnetic activity, and it is therefore interesting to explore correlations with other activity proxies, such as chromospheric activity, and magnetic flux density.
Analogous to coronal  X-ray activity, chromospheric activity also exhibits a large range for stars with similar surface gravity and effective temperature (saturation effects for chromospheric emission are discussed in sec.2.\ref{ssec:1.1} below). 

The X-ray flux shows a good correlation with chromospheric activity indicators based on \caii\ HK core emission, and with the magnetic flux density \citep{schrijver92}. These flux-flux relationships hold for stars (early-F to mid-M type; main sequence, subgiants; and giants down to mid-K), for the Sun as a whole, and for specific solar features extending over 2-4 orders of magnitude, depending on the diagnostics used \citep[see][and references therein]{schrijver08}. \cite{schrijver08} parametrize the relation between magnetic flux density, \mfd, and the radiative diagnostics of activity, $F_i$, as power-laws: $F_i = a_i$ \mfd$^{b_{i}}$.  They find that the exponent $b_{i}$ monotonically increases with the temperature of formation of the radiative diagnostics, ranging from $\sim 0.5$ for chromospheric emission in \caii\ and \mgii, to $\sim 0.75$ for the transition region emission in e.g., \siiv\ and \civ, to $\sim 1$ for the coronal X-ray emission \citep[see also][]{ayres95}.

The correlation of X-ray emission and magnetic flux has been further explored by \cite{pevtsov03} who find a roughly linear relationship between \lx\ and the total unsigned magnetic flux (\lx $\propto \Phi^{1.13 \pm 0.05}$), for  spatially resolved solar features (e.g., active regions, X-ray bright points, quiet Sun regions) and active stars (dwarfs and pre-main sequence), over $\sim 12$ orders of magnitude \citep[see also][]{fisher98}.  These results extend beyond stars for which the coronal filling factor has already reached close to unity (i.e., stars that are completely covered in active regions; see sec.\ref{sec:1}.\ref{ssec:1.2}), indicating that the increase in X-ray emission with magnetic flux
is likely associated with an increase in magnetic field strength and not simply with the integrated surface area of emitting regions.
\cite{pevtsov03} interpret their result as being a consequence of a universal relationship between magnetic flux and power dissipated through coronal heating, and suggest this might imply the presence of one dominant coronal heating mechanism for all solar coronal features and stellar coronae (see section \ref{sec:4} for a brief discussion).

Chromospheric observations of late-type stars also reveal evidence of a ``basal'' emission component that appears unrelated to the stellar magnetic activity, covering the entire surface of magnetically inactive stars \citep[e.g.,][]{schrijver89b,schrijver95}. This {\em basal activity} has been interpreted as due to acoustic heating \citep[see][]{schrijver95}, however, \cite{bercik05} indicate that a turbulent dynamo, from non-rotating plasma, represents a viable alternative to acoustic heating models, and can reproduce the observed basal coronal (X-ray) and chromospheric (\mgii) flux. \cite{cuntz99} have demonstrated that MHD wave heating models using direct stellar magnetic measurements can successfully heat the chromospheres over a wide range of activity.

The close similarity of the phenomenology of solar and stellar activity constitutes the basis of the ``solar-stellar connection''  \citep[e.g.,][]{peres00,peres04}. The challenge of the solar-stellar connection is to investigate the limitations of the solar analogy. In fact, several significant differences in solar and stellar activity phenomena are also observed, especially going towards the extreme of activity. Active stars, for example, tend to have large polar spots \citep[e.g.,][]{berdyugina98,berdyugina00,hussain07}, at odds with less active stars like the Sun.  Also, the high-level of X-ray emission of active coronae, several orders of magnitude larger than that of the Sun, cannot be simply reproduced by scaling up the solar corona, i.e., covering the entire stellar surface with solar active regions \citep[e.g.,][]{drake00}. We now know that active stellar coronae differ in at least some aspects from the solar corona, including, for instance, significantly higher degrees of flaring activity, and higher coronal plasma densities (see sec.\ref{sec:1}.\ref{ssec:1.2}).

\subsection{Activity-rotation relationship}
\label{ssec:1.1}

Since observations of stars c.1980 with the X-ray observatory \einstein\ \citep{giacconi79}, it became clear that rotation was the most important stellar parameter determining the observed activity level  \citep{pallavicini81}.  There is some observed residual dependence on mass of the saturation threshold, and it was discovered that the rotation--activity relationship becomes tighter when using instead of rotation period a Rossby number \rossn $=$\prot$/\tau$, where $\tau$ is the mass-dependent convective turnover time  \citep{Noyes84,pizzolato03}.
\cite{wright11} analysed the  X-ray activity and rotation relation for a large sample including 824 late-type stars with accurately determined X-ray luminosity, \lx, and \prot. 
The relation between  \rx\ and \rossn\  reveals three different activity  regimes: unsaturated (\rossn $>$ \rossnsat $= 0.13$), saturated (\rossn $<$ \rossnsat, and $\log($ \rx $)\sim -3.13$), supersaturated (\rossn $\lesssim 0.01$; \citealt{wright11}).

\begin{figure}[!ht]
\centering
\scalebox{5.25}{\includegraphics{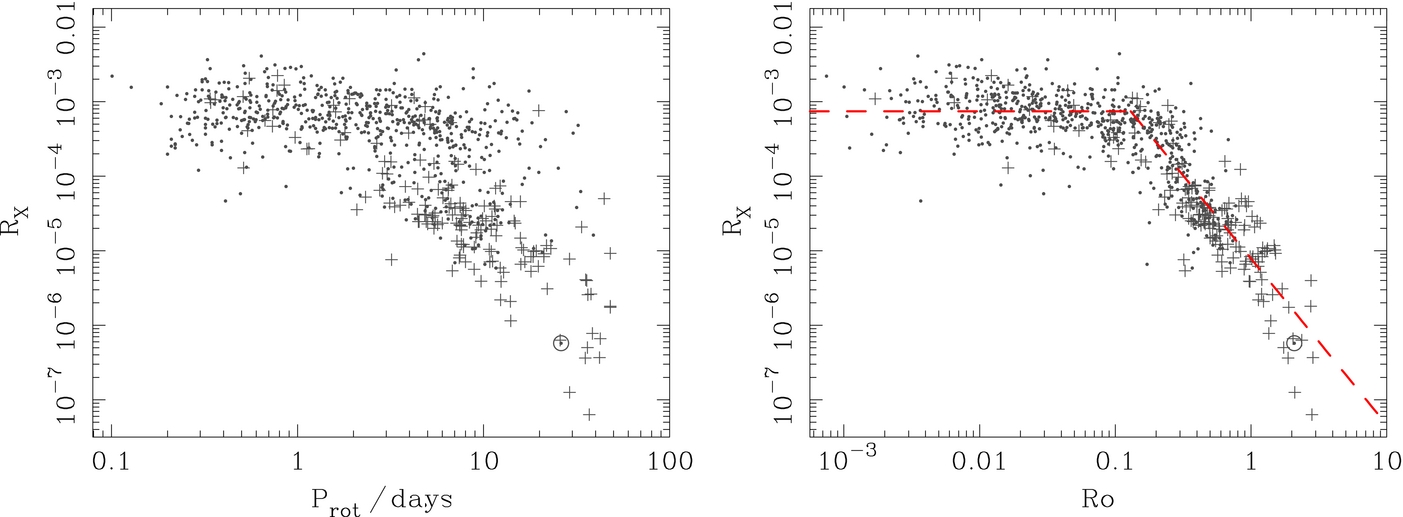}}
\caption{Relationship between \rx\ and \prot\ ({\em left}), and \rx\ and \rossn\ ({\em right}); from \cite{wright11}. For \rossn\ $\gtrsim 0.1$ the stellar activity scales  with \rossn. For \rossn\ below this threshold value the activity level is essentially constant at a saturation value of $\log ($ \rx $)\sim -3$. }
\label{fig:rotation}
\end{figure}

The relation between rotation and convection parameters and
non-thermal losses from coronal and chromospheric activity provided
strong evidence that the latter quantities can be used as proxy
indicators of the efficacy of stellar dynamos
\citep{rosner80,Noyes84}.  Dynamo generation of magnetic fields is
commonly parameterized by a dynamo number ($N_D$) which is essentially the
ratio between magnetic field generation and diffusion.  The Rossby
number was shown by \citet{Noyes84} to be related to the dynamo number
for an $\alpha$-$\omega$ dynamo distributed in the convection zone
according to the relation $N_D=R_0^{-2}$.  However,
\citet{montesinos01} have pointed out that in the currently
favored interface dynamo acting at the tachocline
\citep[e.g.,][]{parker93,charbonneau97}, the relation between the
dynamo number and the Rossby number is not quite so simple. They
showed that the dynamo number does not depend on the Rossby number
alone, but also on a term including the differential rotation across
the tachocline, $\Delta\Omega$.

In the unsaturated regime, the slope of the \rx\ vs.\ \rossn\
distribution then has potentially interesting implications for dynamo theory.  The
steep value of the slope appears to be inconsistent with the
distributed dynamo but can be more easily accomodated by an interface
dynamo acting at the tachocline.  For stars of fairly similar spectral
type, \cite{wright11} reduced the \citet{montesinos01} dynamo
number formulation to $N_D\propto \Delta\Omega/(\Omega R_0^2)$.  If
differential rotation driving the dynamo is proportional to the
rotation rate, \rx\ should vary as the square of the rotation rate
(and similarly, with $1/R_{0}^2$), as was originally observed by
\cite{pallavicini81}.  \citet{reiners14} note that the approximate
dependence of X-ray luminosity on P$_{rot}^{-1/2}$ for the unsaturated
regime indicates that the convective turnover time scales
approximately proportionally to $L_{bol}^{-1/2}$.  However,
\cite{wright11} found instead a best-fit slope of 2.7.  While this is
somewhat dependent on the stellar sample, the observed slope implies a
decrease of differential rotation ($\Delta \Omega \propto
\Omega^{2/3}$) as the star spins down.  This relation is consistent
with those derived from the limited differential rotation data for
single dwarfs ($\Delta \Omega \sim \Omega^{0.7}$, \citealt{donahue96};
$\Delta \Omega \sim \Omega^{1.0}$, \citealt{saar11}).  However, it
becomes problematic in the light of recent measurements of
differential rotation based on high-quality {\it Kepler}
photometry. Those data instead indicate $\Delta\Omega$ to be
essentially constant over a fairly wide range of rotation rate
$\Omega$ \citep[e.g.][]{kuker11,reinhold13}.  Either the
interpretation of the {\it Kepler} photometry is incorrect, or the
elementary dynamo theory will need to be revised to explain the lower
activity end of the rotation--activity relation.

We point out that the observations here are of latitudinal
differential rotation, while the component most relevant for dynamo
action is the radial component.  The two are connected, and likely
scale together, but their relationship is non-trivial (e.g., radial
$\Delta\Omega$ is roughly constant on cones in the Sun, but is
expected to be constant on cylinders in faster rotators).
Astroseismology gives some hope of resolving this issue, at least
for evolved stars \citep{deheuvels14}; radial $\Delta
\Omega$ measurements in dwarfs may still be very challenging 
\citep[e.g.,][]{lund14}.

The saturation value of \rx\ above which faster rotation results in no further increase in \lx, appears independent of spectral type. 
The causes of the observed saturation of X-ray stellar activity are not well established, though several candidates have been identified. The main underlying question to address is whether the saturation of activity is caused by a saturation of the dynamo efficiency \citep[e.g.,][]{gilman83}, or is rather a separate consequence of the fast rotation, such as for instance saturation of the coronal filling factor \citep[e.g.,][]{vilhu84} or centrifugal stripping of the corona \citep[e.g.,][]{jardine99}.
The lack of observed saturation in some chromospheric activity indicators (e.g., \mgii; \citealt{cardini07}) and the saturation of others (e.g., \caii; \citealt{mamajek08,marsden09}) at a different \rossnsat\ value, are in contrast with the idea that the dynamo itself saturates. Analogously, observations of rotational modulation of X-ray emission \citep[e.g.,][]{marino03} and small coronal filling factors \citep[e.g.,][]{testa04b} in saturated stars, indicates that saturation is likely not due to a saturation of the filling factor. The transition between saturated and unsaturated regime could instead be due to a change of regime of the underlying dynamo mechanism from a convective (C) dynamo in fast rotators, where the stellar core and envelope are essentially decoupled, to an interface (I) dynamo for the slower rotators \citep{Barnes03,barnes03a}.  The similar shape, in the mass-age space, of the unsaturated-saturated X-ray activity regimes, and of the C $\rightarrow$ I dynamo (see Fig.~4 of \citealt{wright11}) supports this hypothesis. As noted by \cite{Barnes03}, late M-type stars ($< 0.25$ \msol) are all in the C sequence.

Recently, \citet{blackman15} explain saturation in terms of the relative timescales for shear and convection: if the shear time-scale is shorter than the convective turnover time, convection becomes unimportant for magnetic field amplification.  Towards later spectral types, stars have longer turnover times and saturate at larger values of \prot\ (see e.g., Fig.~6 and 7 of \citealt{wright11}).  This makes sense in terms of the toroidal shear rate being roughly proportional to rotation rate, but again is problematic if the rate of shear is constant for different rotation rates. 
 
Some extremely fast rotators show a decrease in X-ray activity below saturation levels \citep{randich96}.
This {\em supersaturation} of the coronal activity is not observed in M-type dwarfs \citep[e.g.,][]{jeffries11}, but only in earlier spectral types.  Indeed, F-type stars do not appear to go through a saturation phase, but instead evolve from a supersaturation phase directly into the unsaturated regime \citep{wright11}.  Supersaturation does not appear to occur for chromospheric activity  \citep{marsden09,jackson10}, challenging the possibility that supersaturation might be due to a change in the underlying dynamo.

Possible theories put forward to explain supersaturation include: (1) ``coronal stripping'' -- coronal loops become unstable due to centrifugal force, when the star is rotating fast enough that the Keplerian corotation radius gets very close to the stellar surface \citep[e.g.,][]{jardine04}; (2) decrease in filling factor due to poleward migration of the active regions, caused by the effect of rotation on the internal radiative transfer in the star - ``gravity darkening'' effect - and strong convective updrafts in the outer convective envelope \citep[e.g.,][]{stepien01}. \cite{wright11} find that their sample is compatible with both these theories, though a stronger correlation appears to hold for the latter.
 
\subsection{The End of the Main-Sequence \label{ssec:1.1b}}

Stars later than about M4 are fully convective and so the interface
dynamo at the base of the tachocline should not be applicable.  There has
been substantial effort aimed at understanding magnetic activity and
coronal heating around this transition to fully-convective stars in
order to probe changes in the underlying dynamo.  There is no obvious
transition in coronal properties or heating efficiency based on
\rx\ as the fully convective limit is traversed
\citep{fleming95,giampapa96}. \cite{drake96b} showed
that quiescent coronal emission from the M7 dwarf VB8 had remained at
a similar level between 1979 and 1994, and argued for a turbulent
dynamo as being the dominant field generation mechanism
\citep{durney93,weiss93}. Magnetic surface imaging shows the
field geometry changes from mainly toroidal and non-axisymmetric
(M0-M3) to predominantly poloidal and axisymmetric (late M;
\citealt{donati08,morin08}), and it is not obvious how this geometry
might be maintained by a purely turbulent dynamo.

A transition in coronal heating efficiency does exist, however.  A
more sensitive {\it Chandra} survey of very low mass stars and brown
dwarfs by \citet{berger10} found that it occurs at masses
significantly below the fully-convective limit, at a spectral type of
about M8-9. The decrease of coronal emission for these late spectral
types is possibly due to increased neutrality in the photosphere,
which can hamper transport of magnetic energy through the stellar
atmosphere \citep{mohanty02}.

\subsection{Flaring activity \label{ssec:1.2}}

As discussed above, ``nanoflares'' are widely considered a viable candidate heating mechanism for stellar coronae. 
The difficulty in detecting these small flares \citep[see e.g.,][]{testa11,testa12b,testa13,testa14} renders it a challenge to establish their importance for the heating of the solar corona. Investigations of solar flares and microflares indicate a power-law distribution of flaring events as a function of energy ($dN/dE \propto E^{- \alpha}$; \citealt[e.g.,][]{lin84}). \cite{hudson91} showed that the index, $\alpha$, of the power-law distribution has fundamental implication on whether flares might significantly contribute to coronal heating: $\alpha > 2$ implies that smaller events dominate the total energy dissipation rate and that coronal heating could be ascribed to smaller flares. Solar flare studies have yielded values of $\alpha$ between $\sim 1.5$ and $\sim 2.7$ \citep[e.g.,][]{crosby93,parnell00,benz02}, leaving the issue unsolved. A similar range of values for $\alpha$ has been determined for stellar observations \citep[e.g.,][]{audard00,kashyap02}. A recent study, refining the analysis of stellar flares, fitting the photon arrival time distribution with a model of stochastically produced flare emission, for a sample of $\sim 50$ stellar spectra obtained with \chandra, shows that for active stars $\alpha$ is often $>2$, possibly implying flare heating (Kashyap et al.\ in preparation). 

The hypothesis of flare heating being significant, if not dominant, in active stars is also supported by studies of stellar radio and X-ray emission. The radio emission from active stars (including RS CVn binaries and FK Com-type coronae) shows a good correlation with their soft X-ray emission (\lx $\propto L_{R}^{0.73 \pm 0.03}$; \citealt[e.g.,][]{benz94}) over 10 orders of magnitude, from solar microflares (though not solar quiescent emission) to the high activity end of stellar coronae. The similar behavior of solar flares and active stellar coronae supports the importance of flare heating for ``quiescent'' emission of active stellar coronae \citep[e.g.,][]{forbrich11}.  The radio/X-ray correlation is however found to break down for spectral types later than $\sim$ M7, due to a substantial decrease of X-ray activity \citep{berger10}. 

The stellar observations described above suggest a scenario in which flare heating might be significant only for active stars. A scenario with different regimes possibly dominated by different heating processes is also suggested by other stellar data. For instance, studies of coronal densities from high-resolution X-ray spectroscopy show that active stellar coronae are characterized by plasma densities significantly higher than in the solar corona and other low activity stars ($\sim 10^{10}$ cm$^{-3}$ at $\log (T$[MK]$) \sim 6.5$ and $\sim 10^{12}$ cm$^{-3}$ at $\log (T$[MK]$) \sim 6.8$), implying remarkably compact hot coronal structures in active stars (e.g., \citealt{testa04b}; similarly compact coronae are also inferred by different diagnostics, such as X-ray optical depth, and modeling of flares and Fe fluorescent emission,  \citealt{testa04a,Testa07a,Testa07b,Testa08L}).
The derived filling factors increase with \fx, at both $\log (T$[MK]$) \sim 6.5$ and $\sim 6.8$. At a surface X-ray flux level corresponding to the \fx\ typical of solar active regions, the filling factor of $\sim 3$MK plasma approaches unity, while the filling factor of the hotter plasma ($\gtrsim 6$MK) shows a change of slope, increasing more steeply with increasing \fx.
The change of slope could suggest a more rapid increase of flaring activity once the stellar surface becomes covered with active regions (filling factor $= 0.1-1$), possibly due to more interactions between active regions \citep[e.g.,][]{testa04b}.

\subsection{Evolution of activity with stellar age \label{ssec:1.3}}

As discussed above (see beginning of sec.~\ref{sec:1}) both solar and stellar studies have found that X-ray emission correlates well with magnetic flux. 
Several stellar studies have investigated the correlation of the average unsigned magnetic field (\bi), from Zeeman broadening (ZB) measurements, with rotation, and found that \bi\ increases with stellar rotation before reaching a saturation level \citep[e.g.,][]{saar96,saar01,reiners09}, similar to other activity indicators (see e.g., sec.\ref{sec:1}.\ref{ssec:1.1}).
We note, however, that since \bi\ $= f B_{\rm I}$ (where $f$ is the intensity-weighted surface filling factor), the saturation of \bi\ could occur in either $f$ or $B_{\rm I}$, or both.
Analysis by \cite{Cranmer11}, however, suggests that the primary rotational dependence comes through $f$.

\begin{figure}[!ht]
\centering
\scalebox{.35}{\includegraphics{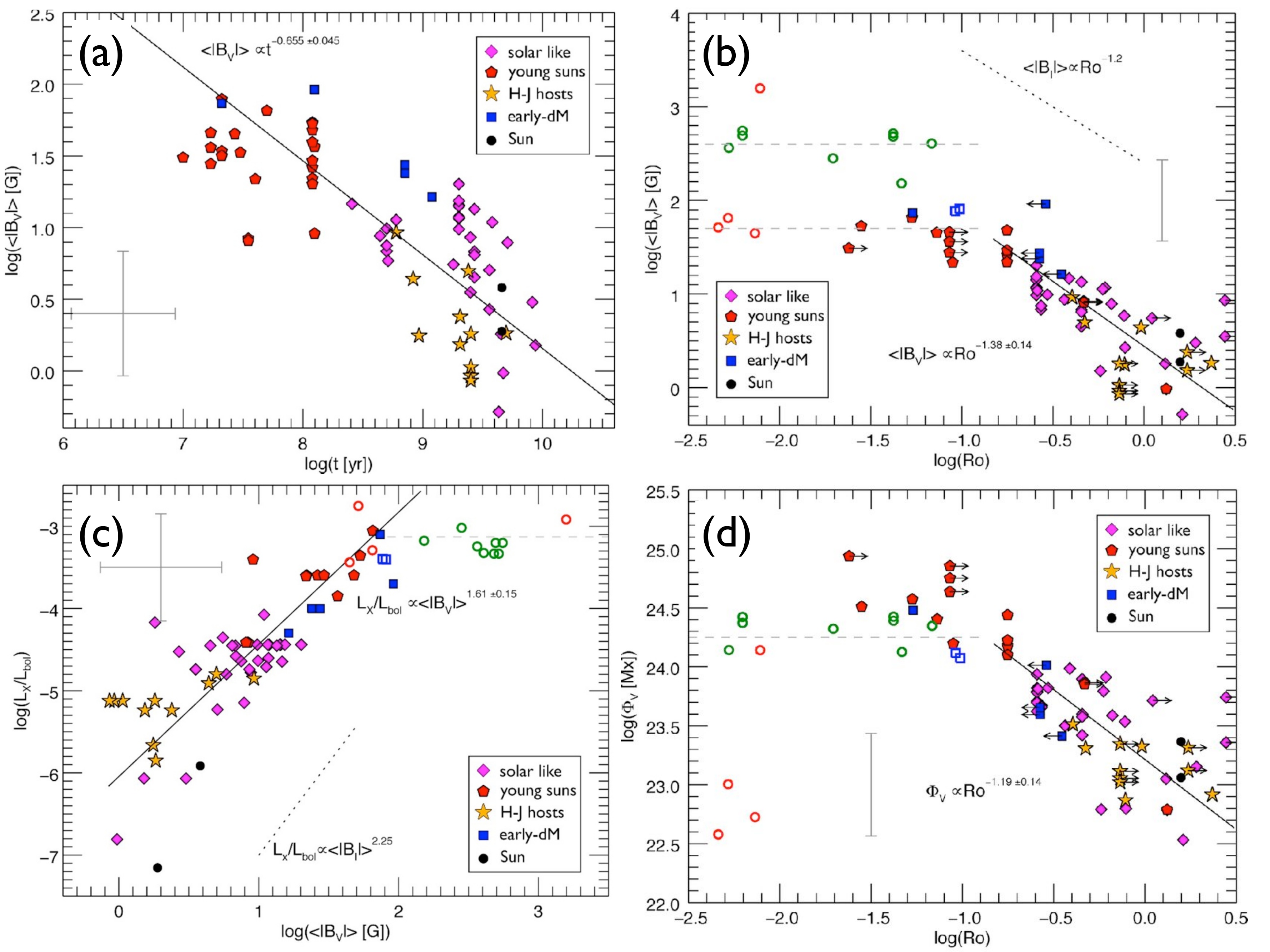}}
\caption{Correlations between average large-scale field strength \bv\ (from Zeeman Doppler imaging), stellar parameters (age, and \rossn), and activity (\rx). {\em (a):} $\log$\bv\ vs.\ $\log$ age, for non-accreting late-type stars. {\em (b):} $\log$\bv\ vs.\ $\log$\rossn; the dotted line is indicative of the correlation, found by \cite{saar01}, between \bi\ (from Zeeman broadening measurements) and \rossn. {\em (c):} $\log ($\rx $)$ vs.\ $\log$\bv; the dotted line shows the slope of the $\log ($\rx $)$ vs.\ \bi\ \citep{saar01,wright11}. {\em (d):} correlation of magnetic flux ($\log \Phi_{\rm V}$) with $\log$\rossn.  All figures are from \cite{vidotto14}.}
\label{fig:Bfield}
\end{figure}

\cite{vidotto14} have recently investigated in detail the relationships between large-scale surface magnetic fields and stellar age, rotation and X-ray emission level for 73 stars ($0.1<$\mstar$<2.0$\msol; 1 Myr $\lesssim$ age $\lesssim$ 10 Gyr). 
From Zeeman Doppler imaging (ZDI) observations they calculate the unsigned surface magnetic field strength (\bv; integrated over the surface of the star) and flux \phiv\ ($= 4 \pi R_{\star}^2$\bv), based on the radial component of the observed surface field (since they are mainly interested in the field associated with the stellar wind), and compare their findings with previous works that used ZB  \citep[e.g.,][]{saar96,saar01}.
The unsigned average large-scale surface field \bv\ decreases with age with a similar dependence (\bv $\propto t^{-0.655 \pm 0.045}$; \citealt{vidotto14}) as the Skumanich law for stellar rotation vs.\ age ($\Omega \propto t^{-0.5}$; \citealt{skumanich72}).
\cite{vidotto14} analyze the relationships between \bv\ and \rossn, as well as \rx\ vs.\ \bv\ and in both cases they find that they are similar (compatible within the uncertainties) to the dependences previously derived for \bi\ \citep[][see Figure~\ref{fig:Bfield}]{saar96,saar01,reiners09}. These results suggest that the ZDI (large scale) and ZB (large and small scale) field measurements are coupled, and that the small and large scale fields share the same dynamo field generation process and they might contribute similarly to the X-ray activity.
A more recent analysis, however, suggests a steeper relation $f \propto$\rossn$^{-b}$ with $2.5 \leq b \leq 3.4$ \citep{Cranmer11}, and hence a similar relationship for $\langle |B_I| \rangle$.  These new slopes are in better agreement with the coronal \rx\ vs.\ \rossn\ \citep{wright11} and the $\Phi_B$ vs.\ \rx\ \citep{pevtsov03} relations, but imply a steeper age dependence than $\langle |B_V| \rangle$.  This is not necessarily a problem; since $\langle |B_V| \rangle$ measures the net flux, it is reasonable to think that as the star becomes more active and crowded with magnetic regions,  $|B_V|$ measurements will be more affected by vector field cancellation and thus rise more slowly with rotation than the average unsigned magnetic field. 

The evolution of the coronal activity of late-type stars with age has been the subject of extensive investigations, and more recently with particular focus on understanding the impact of stellar activity on the evolution of planetary atmospheres \citep[e.g.,][]{ribas05,sanz11}.
Studies of X-ray/EUV/UV emission for a sample of solar analogs at different ages ($\sim 0.1-6.7$~Gyr) revealed that the high-energy stellar emission declines following a power-law relationship with age, and the decline is faster at higher energies (i.e., the power-law slope decreases monotonically from the X-ray to the UV band \citep{ribas05}.
The decline of coronal activity with age is accompanied by lower coronal temperatures and decreased flaring rates \citep{telleschi05}.
\cite{stelzer13} have recenty studied the chromospheric, transition region and coronal emission (by analysing H$\alpha$, UV and X-ray data) of 159 M dwarfs (M0-M3) within 10~pc from the Sun, sampling a wide range of stellar ages, from $\sim 0.01$~Gyr to $\sim 3$~Gyr. Their findings show that the evolution of chromospheric and coronal activity of M dwarfs is similar to higher mass solar analogs, with the X-ray activity levels declining faster with age than the chromospheric UV emission. 
The faster decline with age of X-ray activity compared with the transition region and chromospheric activity can be interpreted in the light of the steeper dependence of the X-ray emission on magnetic flux density, as discussed above at the beginning of sec.~\ref{sec:1} \citep[e.g.,][]{schrijver08}.

 \section{Activity cycles and grand minima} 
\label{sec:2}

The activity of the Sun is known to undergo cyclic variations with an 11-yr period (22-yr period for magnetic activity), as manifest in several activity indicators, including the number and surface area of sunspots, total solar irradiance, chromospheric and coronal activity, and magnetic morphology \citep[e.g.,][]{hudson88,hathaway10}.  
The first evidence of magnetic activity cycles in solar-like stars has been provided by the extensive chromospheric surveys carried out through Mount Wilson \caii\ HK observations \citep{wilson78}, which showed that activity cycles are a common occurrence in late-type stars ($\sim 60$\% of the Mount Wilson main-sequence F- to M-type stars have cyclic chromospheric activity with periods up to $\sim 26$ yrs; \citealt{baliunas95}). Rapid rotators are generally characterized by shorter cycles \citep{saar99,olah09}, but otherwise clear correlations between activity cycles and stellar parameters have not been found.

While cyclic chromospheric activity has been observed for a large number of stars, the detection of X-ray stellar activity cycles is significantly more challenging due to the intrinsic high variability of X-ray emission exhibited by late-type stars and the difficulty of continuous X-ray monitoring.  
To date, convincing X-ray cycles have been seen in only a handful of stars, such as for example 61 Cyg A (K5V; \citealt{hempelmann06}), HD 81809 (G5V; \citealt{favata08}), and $\alpha$ Cen B (K1V; \citealt{ayres09,dewarf10}).   These are all stars of relatively low magnetic activity levels.  At high activity, the evidence to date suggests activity cycles are not present, at least in X-rays.  The eclipsing RS~CVn-type active binary AR~Lac has been observed by all X-ray satellites since the {\it Einstein} observatory in 1980, and data obtained over the past 33 years indicate a remarkable constancy of X-ray luminosity with variations outside of obvious flares typically being about 15\%\ and with a maximum of 45\%\ \citep{drake14}.
In this context, it might be expected that cyclic activity sets in at some activity level (age), as a star spins down. Recently, a short coronal and chromospheric cycle ($\sim 1.6$~yr) has been detected in the $\sim 600$~Myr old solar-like star, $\iota$~Hor (F8V; \citealt{sanz13}). This is the youngest and most active star for which a cycle in chromospheric and coronal emission has been detected so far, and is therefore potentially important to constrain the dynamo properties in young solar-like stars and the onset of activity cycles.  The boundaries between saturated and unsaturated X-ray emission delineated by \citet{wright11} indicate $\iota$~Hor became unsaturated about 400~Myr ago; searching for cycles down to ages of 200~Myrs or so would be of great interest for probing the transition from saturated magnetic activity.

\begin{figure}[!ht]
\centering
\scalebox{.45}{\includegraphics{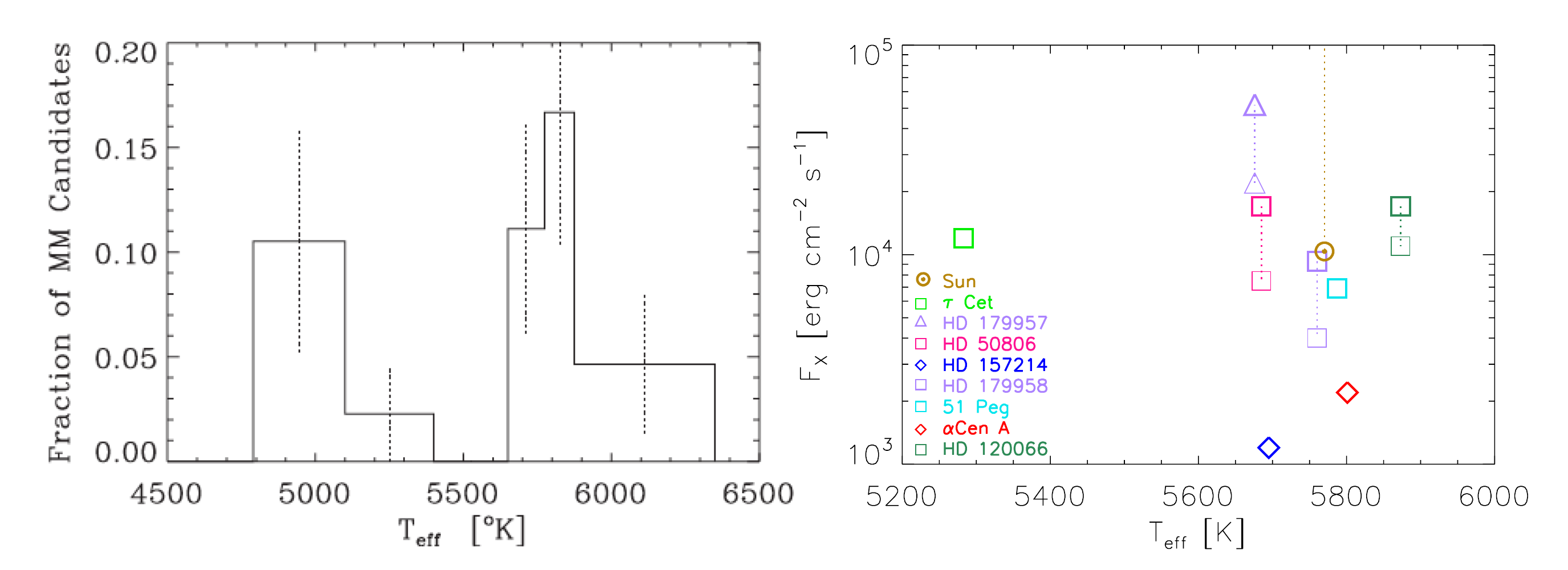}}
\caption{{\em Left}: Distribution of Maunder minima candidates as a function of \teff\ (data from \citealt{saar12}). 
  {\em Right}: X-ray surface flux $F_{X}$ for a few candidates stars in magnetic grand minima and comparison stars (see \citealt{saar12} for details). Squares are MGM candidates, diamonds are stars in temporary minima, and the triangle is the non-MGM companion to the MGM candidate HD 179958. For stars we recently observed with the \chandra\ X-ray observatory \citep{saar12} vertical dots connect fluxes calculated for a single assumed coronal temperature ($T_{\rm X} = 0.1$ keV) or the best fit $T_{\rm X}$ (heavy solid).}
\label{fig:mgm}
\end{figure}

An interesting aspect of the long term variability of stellar activity is the phenomenon of magnetic grand minima (MGM). For the Sun, long term records of sunspot numbers revealed the occurrence of occasional grand minima where some aspects of solar activity (such as sunspot number) have temporarily deviated from cyclic behavior, though other indicators (e.g., $^{10}$Be abundance in ice cores) show a persistence of the magnetic cycle through MGM \citep{beer98}. The relatively short baseline of temporal monitoring of magnetic activity in stars hampers detection of MGM in stars. 
MGM have previously been identified as low-activity stars with chromospheric activity indicator \rhk\ $= (F_{HK}-F_{\rm phot})/F_{bol}$ ($F_{HK}$ is the raw calibrated HK core flux, $F_{\rm phot}$ is the photospheric component of the HK core flux, and $F_{bol}$ is the bolometric flux; \citealt{Noyes84}) below a certain threshold ($\log $ \rhk $\leq 5.1$; \citealt{henry96}). However, \cite{wright04} noted the influence of gravity and metallicity on \rhk, complicating the definition of activity minima based on an \rhk\ threshold. 

We recently refined the MGM candidate definition criteria, in particular to remove metallicity effects, and explored their X-ray emission properties in comparison with low-activity stars \citep{saar12}. 
Taking into account the metallicity, a lower boundary for $\log$\rhk\ is found as a function of metallicity (\rhk $\sim -0.213$ [M/H] $-5.125$) for dwarfs \citep{saar12}.  We therefore defined MGM candidates as dwarf stars (as determined from spectroscopically derived \teff\ and gravity) close to this \rhk\ lower boundary ($\log$ \rhk\ $- \log$ \rhk$_{\rm min} \lesssim 0.054$), and with low chromospheric variability ($\sigma_S/S_{\rm HK} \lesssim 2$\%, where $S_{\rm HK}$ is the chromospheric activity index defined by \citealt{baliunas95}) over timescales of at least 4 yrs \citep{saar12}. We note that these are likely somewhat conservative criteria, as, for instance, higher activity stars, if they experience MGM events, could have MGM \rhk\ values larger than stars at the lowest activity levels. Assuming these criteria and using the sample of \cite{wright+04}, we find that about 7\% of dwarfs are MGM candidates, with no evident dependence on metallicity. The distribution of the MGM candidates as a function of \teff\ peaks near the solar \teff\ (see left panel of Fig.~\ref{fig:mgm}). 

We recently carried out a campaign for X-ray monitoring of a few of these MGM candidates and the derived X-ray fluxes are compared to other low-activity stars in  Fig.~\ref{fig:mgm}. The MGM candidates have X-ray emission similar to the Sun at normal magnetic cycle minima, though their coronae seem cooler than the solar corona at minimum, perhaps suggesting a larger contribution from coronal holes rather than quiet sun-like regions in these extremely low-activity coronae. The X-ray fluxes of these low-activity stars can be compared with typical \fx\ of the solar quietest coronal features, such as quiet sun and coronal holes.  For the quiet Sun \fx\ $\gtrsim 2 \times 10^{4}$ erg cm$^{-2}$ s$^{-1}$ \citep{chamberlin09,linsky14}, while for coronal holes \fx\ is approximately in the range $5-10 \times 10^{3}$ erg cm$^{-2}$ s$^{-1}$ (e.g., \citealt{schmitt12} estimated \fx\ $\sim 10^{4}$ erg cm$^{-2}$ s$^{-1}$, while we estimated \fx\ $\sim 7 \times 10^{3}$ erg cm$^{-2}$ s$^{-1}$ by using temperature and emission measure estimates from \cite{kano08} and synthesizing the emission using CHIANTI; \citealt{chianti,chianti7}). 
Some of these low-activity stars can reach \fx\ values possibly even lower than the \fx\ of solar coronal holes cited above, but these appear to be unusually cool and weak (normal) cycle minima, rather than extended ``grand minima'' states \citep{saar12}.
We note, however, that direct comparisons between solar and stellar values are often not straightforward, for several reasons including (a) different observing passbands of solar and stellar instruments, and (b) significant uncertainties in the \fx\ estimates above, reflecting uncertainties in the derivations of temperature and emission measure of coronal holes, e.g., due to scattering and contribution of overlying QS plasma, especially at the CH boundary (as well as other assumptions, e.g., on coronal abundances).

Solar and stellar observations of activity cycles and cycle grand minima are crucial to constrain  dynamo models, and investigate the activity conditions and other properties of MGM, their underlying causes and transient nature (i.e., what dynamo properties determine the onset and duration of MGM).  Since the Sun is currently not in such a state, stars offer the only avenue of observational attack.

  \section{Abundance anomalies} 
\label{sec:3}

Early spectroscopic observations of the solar corona revealed a chemical composition different from the underlying photosphere, in particular with a dependence on the element's first ionization potential (FIP): low FIP (e.g., Mg, Fe, Si) elements appear 2-4 times more abundant in the corona than in the photosphere, compared to high FIP elements (e.g., C, N, O; \citealt[e.g.,][]{meyer85,feldman92}).
X-ray and EUV high-resolution spectroscopy showed that other stars also exhibit fractionation effects in their coronal composition compared with photospheric abundances (see e.g., review by \citealt{testa10ssrv} and references therein), and a {\em FIP effect} similar to the one observed on the Sun was observed for some low activity stars, such as e.g., $\alpha$~Cen \citep[e.g.,][]{drake97}.
The chemical composition of the coronal plasma appears to depend on the stellar activity level, with active stars characterized by an {\em inverse} FIP effect, i.e., with coronal enhancement of high FIP elements \citep[e.g.,][]{brinkman01,garcia09,testa10ssrv}.
The coronal abundance patterns and their possible dependence on stellar parameters and activity levels are interesting in the context of coronal heating.  The fractionation is thought to occur during transport to there corona from lower altitudes, as species transition from being neutral to ionized.  As such, abundance anomalies offer the potential to probe the physical processes behind this transport, which are also likely related to the mechanism(s) of stellar atmospheric heating. 

Recently, additional studies have suggested that, whereas for high activity stars the coronal abundances depend on activity level, for low to moderate activity (main-sequence) stars ($\log$ \lx $< 29.1$) they might have a stronger dependence on spectral type \citep{wood10,wood12,wood13}.  A change in abundance patterns between high activity and low/intermediate activity levels might reflect a transition between different heating regimes, as discussed in sec.\ref{sec:1}\ref{ssec:1.2} above, in the context of flaring activity.

\begin{figure}[t]
\hspace{-2.5cm}\scalebox{.63}{\includegraphics{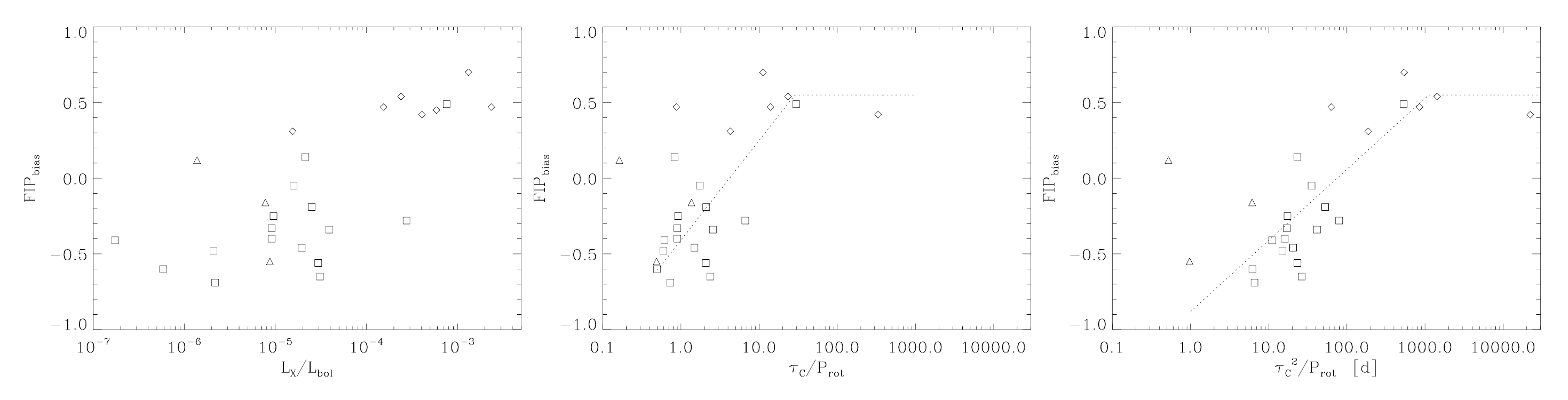}}
\caption{For the source sample of \cite{wood12} and \cite{wood13}, we plot FIP bias vs.\ \rx\ ({\em left}), vs.\ 1/\rossn\ ({\em center}), and vs.\ $\tau_c$/\rossn\ ({\em right}).}
\vspace{-0.15cm}\label{fig:fip_lxlbol}
\end{figure}

To further explore the effect of activity level on coronal abundances we use the measurements by \cite{wood12} and \cite{wood13} and plot their measured FIP bias as a function of the activity index \rx\, and as a function of 1/\rossn\ (left and center panel of Figure~\ref{fig:fip_lxlbol} respectively). These plots suggests that a dependence of chemical fractionation on stellar activity is present also for low to moderate activity stars, likely together with a spectral type dependence. This is in agreement with the apparent activity dependence of Ne/O at low activity levels \citep{robrade08} and in solar active regions of different plasma temperature \citep{drake11,landi15}.  High activity coronae do not exhibit such dependence but instead appear to show a ``saturated'' coronal Ne/O \citep{draketesta05}, as also discussed by \cite{wood12}. 
Figure~\ref{fig:fip_lxlbol} (center panel) shows  a general increase of FIP bias with $\log$\rossn\ up to saturation at $1/$\rossn\ $\sim 10$, with some outliers.  Investigation of the outliers suggests an additional mass, and specifically, convection zone-related term might further improve the relationship. The right panel of Figure~\ref{fig:fip_lxlbol}  explores an additional dependence on $\tau$ (i.e., $\tau/$\rossn); the result shows reduced scatter, with only one star (Procyon) significantly discrepant.
We note that the FIP saturation point ($1/$\rossn\ $\sim 10$) approximately corresponds to the saturation threshold of coronal emission (Fig.~\ref{fig:rotation}), suggesting a possible connection between the phenomena: the maximum FIP bias might be limited by the maximum coronal activity possible.

Efforts have been made in searching for a unified model to explain chemical fractionation in stellar coronae, in particular investigating the effect of ponderomotive forces associated with the propagation of Alfv$\acute{\rm e}$n waves through (or reflection from) the chromosphere \citep[e.g.,][]{laming04,laming09,laming12,wood13}.
While these studies provide a promising start, further progress is needed, both on the observational side (e.g., a more thorough disentangling of dependence on different stellar parameters, improved determination of abundance anomalies in different solar coronal features), and on the theoretical modeling, in order to exploit the potential of coronal abundances as diagnostic of coronal heating processes.

\section{Discussion \& Open Questions}
\label{sec:4}

We have attempted a very brief summary of some recent developments in the study of stellar coronal activity within a broad context of coronal heating.  While the general picture of stellar coronae has been greatly clarified over the last decade or so, as stellar studies become more detailed and instrumentation more powerful, almost as many new questions are raised by the new data as are answered.  Here, we outline what we consider to be some of the more important outstanding issues, and the observational and theoretical avenues we need to take to begin to address them. 

\begin{enumerate}

\item {\it What are the respective roles of interface dynamos and purely convectively-driven dynamos in stars?}  The realization that magnetic dynamos ultimately give rise to coronal heating raised the question of how the dynamo behaves in stars without the means for rotational shear at the base of the convection zone---in fully convective T~Tauri stars or very late M dwarfs for example.  That there is no precipitate decline or change in coronal properties between fully convective stars and stars with convectively stable cores is perhaps hinting that the tachocline is not such a vital part of the process.  Dynamo theory for non-solar cases would bear revisiting, especially in the light of differential rotation measurements (below).  

\item {\it How do stellar activity and magnetic dynamos map to the growing number of observations of stellar differential rotation?} The simplicity of the relation between unsaturated X-ray emission
and the square of the Rossby number is beguiling, but elementary
dynamo models depend on radial differential rotation scaling with
rotation rate.  Yet both theory and {\it some} observations (of
admittedly, as noted in sec.\ref{sec:1}.\ref{ssec:1.1} above, the {\it latitudinal} shear component)
are converging on shear with only a very weak rotation rate dependence.
Rotational shear must play a role in magnetic field generation, but
in dynamo models a constant shear implies either a stronger rotational
dependence of the convective term ($\alpha$-effect), which seems
unlikely, or else rotation plays a role not yet considered.  The
logical step would be, ironically, backwards, toward the idea in
which the dynamo is more distributed through the convection zone
than seated at the tachocline.  One problem is that the large {\it
Kepler} sample of differential rotation measurements has not been
studied to any extent in X-rays, or with a technique that provides
an unambiguous diagnostic of magnetic flux generation.  They are
also less well characterized in terms of their binarity or evolutionary
status: \cite{saar09,saar11} argue that binary and evolved stsrs
should be considered separately, as they appear to have different,
reduced $\Delta\Omega$ dependences.  X-ray luminosities combined
with differential rotation measurements for well-characterized
stars will provide the necessary data to move forward.  The {\it
Kepler} K2 surveys will include well-studied stars in a variety of
clusters (e.g., M35, M44, the Pleiades, Hyades and Praesepe) that
have detailed X-ray data, and could provide an interesting step
forward.

\item {\it How does flaring behavior depend on spectral type and stellar luminosity class?} We tend to use the term ``flare" somewhat loosely, and while the general properties of flares appear similar, their detailed  characteristics are probably dependent on the magnetic field morphology---loop sizes, active region structure and field strength---that in turn probably depends on underlying stellar properties such as convective vigor, differential rotation and surface gravity.   Stars like the X-ray bright evolved Capella binary (the brightest stellar X-ray source in the sky), and the nearby F6 sub-giant Procyon are notorious for showing little or no obvious X-ray flaring activity \citep[e.g.][]{testa12}.  This is telling us something about how their surface magnetic fields are configured and excited (or not). The energy dissipated by large solar flares is related to the magnetic potential of parent active regions \citep[e.g.][]{schrijver12}.  A systematic study of flare energies and their distributions on stars over a range of spectral types could provide us with a picture of surface magnetic potential energy and how it is shaped by stellar characteristics.   
Another aspect to investigate is whether the energy partition in stellar flares differs from the solar case. The argument above provides some rationale for flares to behave differently on different types of stars.  {\it Chandra} and {\it XMM-Newton} have succeeded in detecting the Neupert effect, indicating the relationship between evaporated and/or heated flare plasma and accelerated particles, but as in the case of the Sun, detection of blueshifts expected from evaporating plasma in loops have proved elusive.  Next generation missions (e.g. {\it ASTRO-H, Athena})  carrying calorimeter-based imaging spectrometers with high spectral resolution at high energies to observe H-like and He-like complexes of abundant metals such as Ca and Fe will be needed for further observational progress.  At higher energies,  only a small handful of giant stellar flares have been studied in hard X-rays, where non-thermal signatures are easier to disentangle from thermal emission.  Current missions with hard X-ray capability are still too insensitive to study stellar flares routinely.  Sensitivity at least an order of magnitude greater than that of {\it NuSTAR} would be needed to make progress.  Since the majority of the energy in solar flares is released in white light or CMEs (see below), simultaneous white light and X-ray observations are needed determine stellar flare total energies.  {\it Kepler} has also provided a rich vein of white light flare data, but again on stars with no X-ray monitoring.   Simultaneous optical and X-ray observations of flares are few and far between.  In this context, X-ray campaigns coordinated with the K2 survey would be very valuable.

\item {\it How is the magnetic energy dissipated in stellar outer atmospheres partitioned between coronal heating, winds and coronal mass ejections?} For the Sun, energy dissipated in the outer atmosphere is divided fairly equally between coronal heating and wind kinetic energy, with coronal mass ejections accounting for 5 to 10\% (depending on cycle phase) of the solar wind mass loss \citep{webb94,priest14}.  Solar-like winds are very difficult to study on even the most nearby stars, and the scant studies to date suggest that stellar wind mass loss does scale with X-ray luminosity up to a certain activity level.  The \citet{wood14} study of the 500 Myr old solar analog $\pi^1$~UMa instead reveals essentially a solar mass loss rate at very high activity levels.  This suggests a fundamentally different outer atmospheric magnetic field structure.  The role of CME's in mass loss on other stars is also very unclear.  Scaling solar flare--CME relations to the flaring rates of stars suggests prohibitively large CME mass loss rates requiring unrealistically high amounts of energy \citep{drake13}.  The implication is that large stellar flares put a much lower fraction of their energy into CMEs.  A much larger survey of stellar winds is needed to better understand their dependence on stellar activity and spectral type.  Unfortunately, the technique, based on Ly$\alpha$ observations, is extremely challenging and it is probably infeasible to expand the existing {\it Hubble Space Telescope} sample beyond a very small number of additional targets.  CMEs on stars cannot presently be observed directly.  Their cumulative effect should be seen in the same observations used to probe stellar winds, however.  The indication of much lower mass loss rates in the most active stars suggest a very interesting change in the way flares work, with the implication that the flare-CME relation is indeed fundamentally different \citep{drake13}.  More observations of stellar mass loss are needed for active stars to make progress; unfortunately this would seem to require new observational techniques, and perhaps new missions.

\item {\it How does coronal heating work near the end of the main-sequence, where stellar photospheres are essentially completely neutral?}  The observation of a break in the ratio of X-ray to bolometric energy output of stars at the very end of the main sequence is very interesting for coronal heating and provides a parameter space that, if we could study it in detail, might teach us much about how different candidate processes work.  The data suggest that the amount of magnetic energy that goes into particle acceleration---electrons at least, that give rise to the observed radio signatures---does not change appreciably going to these very cool dwarfs, but that plasma heating is efficiently quenched.  It is expected that heating should decline as the atmosphere becomes more neutral due to large resistivities and efficient field diffusion, and consequently less effective driving of magnetic stresses.  But why should particle acceleration remain efficient, and isolated flaring activity be present?   Much longer and deeper observations of nearby objects are needed to better understand their X-ray activity, while a wider survey covering a larger number of object would probe coronal heating over a wider range of parameters and rotation rates.  Such studies are partially within reach of current missions, although would require expensive commitments of time. 

\item {\it How do coronal abundance anomalies correlate with underlying stellar properties and with coronal properties such as magnetic scale height, plasma temperature and flaring rate?}  Theoretical basis for understanding the FIP Effect has historically not been lacking---it has been easy to point to suitable regions in the chromosphere where the separation is likely to take place and take advantage of those conditions to effect it---but models have tended to fall short, and the avalanche of new stellar data at the turn of the millennium from \chandra\ and \xmm\ tended to fall outside of the scenarios tailored for the Sun.  The ponderomotive force-based model proposed by Laming and collaborators offers an appealing solution to the problem that can in principle explain both enhancements and decrements in species under different conditions.  While the observational database is presently fairly large, the solar case illustrates that different regions of the same star can exhibit fairly dramatic differences in chemical composition.  Stars are likely to do the same things, and what we observe is an emissivity-weighted melange of that behavior.  Abundance estimates are often fairly uncertain and carry considerable scatter, though they are also likely to be time-dependent according to flares and dominant active region evolution and rotational modulation.  Future X-ray instruments with larger effective areas will be able to probe the time-dependence of abundance anomalies, and begin to separate the ensemble behavior into its constituent parts.  Long commitments of observing time on present day instruments could also tease out secular trends, but will be difficult to justify to allocation committees.  Theoretical advances through more detailed and realistic modeling can only help the arguments to secure the observational data.  

\end{enumerate}

We look forward to the next few years to see how the many facets of the coronal heating problem evolve, and to the new array of questions and problems that both theoretical and observational progress are inevitably going to uncover!

\begin{acknowledgements} 
The authors would like to thank N.~Wright and A.~Vidotto for giving permission to reproduce figures from their papers. We would also like to thank Fabio Reale, Karel Schrijver, Vinay Kashyap, and Cecilia Garraffo for useful discussions, and the anonymous referees for insightful comments which helped us to improve the paper.
This work was supported by NASA under grant NNX10AF29G, contract NNM07AB07C to the Smithsonian Astrophysical Observatory, Chandra grant GO1-12036X, and contract SP02H1701R from Lockheed-Martin to SAO.
\end{acknowledgements} 
 
\bibliographystyle{rspublicnat}

\end{document}